# Comment on Carlo Rovelli's "An argument against the realistic interpretation of the wave function"  - arxiv:1508.06895v2

H. Dieter Zeh

University of Heidelberg – www.zeh-hd.de

In a new paper,[1] Carlo Rovelli argues that the time reversal asymmetry of quantum measurements be in conflict with a realistic interpretation of quantum states. Let me point out that his conclusion is essentially based on an inappropriate formulation of quantum measurements.

In its first part, the paper refers to the $T$ symmetry of the time-dependent Schrödinger equation and that of transition probabilities according to the equality of $|<a|b>|$ and $|<b|a>|$.[2] Traditionally and justifiably, the presence of these two quite different dynamical laws *for the wave function of the measured system by itself* (with values of the observables represented by their eigenfunctions) has indeed been regarded as inconsistent with an ontic interpretation. Instead of recalling this well known argument, the author uses this (simplified) description of a measurement in order to compare the "textbook prescription" of a time-reversed measurement (of the first kind), which requires that the eigenfunction of an observable $B$, say, always applies immediately *after* a corresponding measurement, with the result of a formal $T$ transformation in which this $B$ measurement would have to affect its past rather than its future! Otherwise, his Equ. (4) would describe two $A$ measurements rather than a $B$ and an $A$ measurement (as assumed). The contrast between these two kinds of time reversal is then regarded as evidence against an ontic interpretation of the wave function in general, while the presence of two different dynamical laws is not further mentioned.

Obviously, the empirical time direction of measurements is induced by the thermodynamical arrow underlying the retarded nature of the apparatus and its registration device. In a *consistent* ontic interpretation, however, the apparatus has to be described by the wave function, too. The asymmetry of the collapse for the combined system according to von Neumann (namely, as eliminating the entanglement previously arising from the measurement interaction) is, of course, also well known. It has even been proposed as being responsible for the thermodynamic and radiation arrows in the universe, while the thermodynamical arrow of the macroscopic apparatus has occasionally been claimed to explain the asymmetric collapse. None of these proposals is sufficient, but none of their arrows is in conflict with realism.

Interaction with the apparatus can quantum mechanically be taken into account with or without a collapse that would supplement the Schrödinger equation. Both descriptions require special initial conditions of negligible initial entanglement for the measurement process. As the collapse is asymmetric by definition, it can resolve the arising problem of superposed pointer positions in a realistic and objective sense, and it might even produce the required initial conditions for subsequent measurements. (As discussed by Roger Penrose long time ago,[3] even this asymmetric collapse can in principle be formally time-reversed by redefining it correspondingly, that is, as acting backwards in time as it would be required in Rovelli's (4). This is true, however, only as long as the apparatus can be shielded against irreversible decoherence[4] – a situation that would describe "virtual" measurements only.) All real (irreversible) measurements – not only the human bookkeepers mentioned in the paper – must exploit past low entropy by using an external arrow of time, and correspondingly affect the combined system.

The author mentions decoherence only for no-collapse interpretations, but without referring to its fundamental irreversibility. Decoherence represents *arising* entanglement of the pointer position with an uncontrollable and unbounded environment. Its arrow of time, which is quite analogous and related to Boltzmann's chaos assumption for classical molecules, clearly requires a *cosmic* initial condition (at the big bang, say) of lacking or small entanglement if the Schrödinger equation holds universally. This would explain the asymmetric branching in time (the "tree" of Rovelli's Figure 1), and it is absolutely compatible with a global ontic wave function that obeys the $T$-symmetric Schrödinger equation[5] (or even with the timeless Wheeler-DeWitt equation that must definitely describe multiple quasi-classical worlds[6]). A $T$ symmetry of cosmic history (rather than that of its dynamical laws only) could then possibly – but need not – be restored on a time scale that is very many orders of magnitude larger than the present age of the universe (its distance in time from the big bang).[7]

I do agree with the author that Bohm trajectories are no more than unphysical "pointers" that indicate in which branch of the wave function *we* happen to live,[8] while all other branches must continue to "exist" according to any no-collapse interpretation. However, I deeply disagree with his proposed description of the real world in terms of stochastic space-time *events*, because all kinds of apparent events have been demonstrated theoretically and/or experimentally to form fast but smooth decoherence processes in accordance with a time-dependent Schrödinger equation that includes the environment. This observed smooth and *T*-asymmetric process can hardly be assumed to occur merely in the eye of the beholder. Thinking in terms of events may indeed be the reason for the author's misinterpretation of his Equ. (4). There is no evidence for any *genuine* fundamental events, which the founders of quantum theory (who gave this theory its name) believed to have discovered, while the branching of the wave function is a dynamical approximation, valid under the mentioned initial conditions. It becomes relevant only for the subjective observer and his "frog's perspective".[9] Any conceivably underlying deeper theory would thus have to explain the superposition principle with its generic and often confirmed consequence of entanglement, but no events.